\documentclass[useAMS,usedcolumn,usegraphicx,usenatbib]{mn2e}

\title[Sensitivity of CO emission to parameter variations]
      {Molecular line intensities as measures of cloud masses - \\ I. Sensitivity of CO emissions to physical parameter variations}
\author[T.~A.~Bell et al.]
{T.~A.~Bell,$^{1}$\thanks{E-mail: tab@star.ucl.ac.uk}
E.~Roueff,$^{2}$
S.~Viti$^{1}$
and D.~A.~Williams$^{1}$\\
$^{1}$Department of Physics \& Astronomy, University College London, Gower Street, London WC1E 6BT\\
$^{2}$LUTH, Observatoire de Paris, Section de Meudon, \break Place Jules Janssen, 92195 Meudon, France}

\begin{document}

\date{Accepted 2006 July 15. Received 2006 July 13; in original form 2006 May 26}

\pagerange{\pageref{firstpage}--\pageref{lastpage}} \pubyear{2006}

\maketitle

\label{firstpage}

\begin{abstract}
A reliable estimate of the molecular gas content in galaxies plays a crucial role in determining their dynamical and star-forming properties. However, H$_2$, the dominant molecular species, is difficult to observe directly, particularly in the regions where most molecular gas is thought to reside. Its mass is therefore commonly inferred by assuming a direct proportionality with the integrated intensity of the $^{12}$CO($J=1\to0$) emission line, using a CO-to-H$_2$ conversion factor, $X$. Although a canonical value for $X$ is used extensively in such estimates, there is increasing evidence, both theoretical and observational, that the conversion factor may vary by over an order of magnitude under conditions different to those of the local neighbourhood. In an effort to understand the influence of changing environmental conditions on the conversion factor, we derive theoretical estimates of $X$ for a wide range of physical parameters using a photon-dominated region (PDR) time-dependent chemical model, benchmarking key results against those of an independent PDR code to ensure reliability. Based on these results, the sensitivity of the $X$ factor to change in each physical parameter is interpreted in terms of the chemistry and physical processes within the cloud. In addition to confirming previous observationally derived trends, we find that the time-dependence of the chemistry, often neglected in such models, has a considerable influence on the value of the conversion factor.
\end{abstract}


\section{Introduction}\label{Introduction}
Accurately measuring the mass of molecular gas in galaxies is of great importance in determining the dynamical properties and star formation efficiency that are critical to models of galaxy evolution. Such measurements also play a vital role in understanding the properties of the interstellar medium in extragalactic environments. Molecular hydrogen (H$_2$) is the dominant molecular species in the Universe and governs the dynamical and chemical evolution of molecular clouds. However, its lack of a dipole moment means that it can only change ro-vibrational state through weak quadrupolar transitions that are only rarely detectable. Alternative means of deriving the mass of molecular clouds have therefore been developed over the years. Many of these techniques make use of a tracer species (a molecule or dust) that is easier to detect and infer the molecular hydrogen content by some proportionality relation.

One of the most commonly used tracers of molecular gas is carbon monoxide (CO). After H$_2$, it is the most abundant molecule in molecular clouds by several orders of magnitude and has been observed in all regions where molecular gas is thought to be present. However, unlike molecular hydrogen, CO readily emits under the conditions typically found in molecular clouds and is easily observed at radio and submillimetre wavelengths with ground-based facilities.

The column density of molecular hydrogen along a line of sight is often related to the integrated intensity of the $^{12}$CO($J=1\to0$) rotational transition line using the CO-to-H$_2$ conversion factor:

\begin{equation}\label{XFactor}
 X = \frac{N(\rmn{H_2})}{\int T_{\rmn{A}}(\rmn{CO})\,\rmn{d}v}\qquad[\rmn{cm^{-2}\,(K\,km\,s^{-1})^{-1}}]
\end{equation}
where $N(\rmn{H_2})$ is the column density of H$_2$ and $T_{\rmn{A}}(\rmn{CO})$ is the antenna temperature of the CO(1--0) line \citep[see, e.g., reviews by][]{Maloney1990,Combes1991,Young1991}.

Observational estimates of $X$ have been derived for nearby regions where other methods for determining molecular cloud masses are available. Values of $X$ were first determined using extinction measurements obtained from star counts to derive $N(\rmn{H_2})$, assuming a constant gas-to-dust relationship, and comparing with observed CO luminosities \citep[e.g.][]{Dickman1978,Frerking1982}. \citet{Scoville1987} and \citet{Solomon1987} derived values of $X$ for a large sample of resolved giant molecular clouds with assigned kinematic distances by calculating their virial masses from CO linewidths. The same approach has recently been applied to clouds within nearby galaxies \citep[e.g.][]{Rosolowsky2003,Rosolowsky2005}. It should be noted that the values derived using this method are based on the assumption that the clouds are virialized. Maps of the diffuse $\gamma$-ray emission produced by interactions between cosmic rays and hydrogen nuclei have been compared to \mbox{H\,{\sc i}} and CO survey data for the Galactic plane to calculate the contribution of H$_2$ nuclei to the $\gamma$-ray emission and so calculate $X$ values along many lines of sight \citep*{Blitz1985,Bhat1986,Richardson1988,Strong1988,Bloemen1989,Strong1996}. The remarkable outcome of these studies is that the value of $X$ seems to be roughly constant for the Galactic molecular clouds that have been observed, and a canonical value is often adopted: $X\approx2\times10^{20}~\rmn{cm^{-2}\,(K\,km\,s^{-1})^{-1}}$ \citep*{Strong1996,Dame2001}. There have been several attempts to explain this apparent invariance, citing ensembles of virialized clouds \citep*{Dickman1986,Solomon1987} and clumpiness of the interstellar medium on scales of $A_V<2$~mag \citep*{Taylor1993}.

Despite the widespread use of this canonical value, there is strong evidence to suggest that $X$ can vary significantly under conditions differing from the local environment. Even within the Milky Way, observationally determined values for the $X$ factor are found to vary with Galactocentric radius \citep*{Digel1990,Sodroski1995,Dahmen1998} and, further afield, estimates for $X$ in the Small and Large Magellanic Clouds are over an order of magnitude higher than the canonical value \citep{Cohen1988,Rubio1991,Israel1997}. Conversion factors up to an order of magnitude lower than the canonical value have also been determined for a number of galaxy nuclei \citep*{Smith1991,Israel2001,Israel2006}. Theoretical studies of the variation of the CO-to-H$_2$ ratio have also suggested that $X$ can be up to an order of magnitude higher or lower than the canonical value under certain conditions \citep[e.g.][]{Sakamoto1996,Sakamoto1999}. \citet{Taylor1993} used a photon-dominated region (PDR) chemical model to perform {\it ab initio} calculations of $X$ and showed its variation for small changes in typical cloud parameters. Work along similar lines has also been conducted by \citet*{LeBourlot1993a}, \citet*{Wolfire1993} and \citet{Kaufman1999}. These previous studies provide the foundation and motivation for this current project.

Since the majority of CO emission occurs in the outer regions of molecular clouds, where the interstellar radiation field falls on the cloud surface, PDR models are most suitable to the theoretical study of the $X$ factor. PDRs arise wherever FUV radiation ($6<h\nu<13.6$ eV) is incident upon the outer edge of a molecular cloud. The radiation drives photochemistry within the cloud and governs the chemical and thermal structure near its surface.

In this paper we examine the influence of physical and chemical conditions on $X$ and attempt to explain its dependence on important physical parameters, such as the density, age, and metallicity of the cloud. In Section~\ref{Models} we describe the PDR models used to calculate and independently verify theoretical estimates of the CO-to-H$_2$ ratio for given physical conditions. Section~\ref{Benchmarking} reviews the independent benchmarking results for important combinations of parameters when using the two PDR models and evaluates the reliability of subsequent, more detailed calculations. The effect of varying physical parameters is then discussed in depth in Section~\ref{Results}, before we summarize our conclusions in Section~\ref{Conclusions}.

\section{PDR models}\label{Models}
\subsection{The {\sevensize\bf UCL\_PDR} code}\label{UCLPDR}
We use the \textsc{ucl\_pdr} time-dependent PDR code \citep*{Papadopoulos2002,Bell2005} to calculate the chemistry, thermal balance and CO(1--0) emission strength within the outer region ($A_V<10$~mag) of a cloud to derive theoretical values of the CO-to-H$_2$ ratio, $X$, for given physical parameters.

The PDR is modelled as a one-dimensional semi-infinite slab, illuminated from one side. The chemistry and thermal balance are calculated self-consistently at each depth point into the slab and at each time-step, producing chemical abundances, emission line strengths and gas temperatures as a function of depth and time.

We adopt a chemical network containing 128 species and over 1700 reactions, including ion-molecule, photoionization and photodissociation reactions. Freeze-out of atoms and molecules onto grains is neglected. The gas is assumed to be initially in atomic form, with the elemental abundances listed in Table~\ref{Abundances} (all fractional abundances in this paper are quoted relative to total hydrogen nuclei). Elemental abundances of all metals are assumed to scale linearly with metallicity ($Z/Z_\odot$). The reaction rates are taken from the {\small UMIST99} database \citep*{LeTeuff2000}, with some modifications introduced as part of a recent PDR benchmarking effort \citep{Rollig2006}. We adopt the grain surface H$_2$ formation rate of \citet{deJong1977}:

\begin{equation}\label{FormationRate}
 R=3\times10^{-18}\sqrt{T}\exp(-T/1000)\qquad[\rmn{cm^3\,s^{-1}}]
\end{equation}
and assume that the rate scales linearly with metallicity.

\begin{table}
 \caption{Elemental abundances used in the {\sc ucl\_pdr} code (relative to total hydrogen nuclei).}
 \label{Abundances}
 \begin{center}
  \begin{tabular}{l r @{$\times$} l l r @{$\times$} l}
  \hline
  He & 7.50 & 10$^{-2}$ & C  & 1.42 & 10$^{-4}$ \\
  N  & 6.50 & 10$^{-5}$ & O  & 3.19 & 10$^{-4}$ \\
  Na & 8.84 & 10$^{-7}$ & Mg & 5.12 & 10$^{-6}$ \\
  Si & 8.21 & 10$^{-7}$ & S  & 1.43 & 10$^{-6}$ \\
  Cl & 1.10 & 10$^{-7}$ & Ca & 5.72 & 10$^{-10}$\\
  Fe & 3.60 & 10$^{-7}$ \\
  \hline
  \end{tabular}
 \end{center}
\end{table}

\begin{table}
 \caption{Standard parameter values adopted for each cloud model.}
 \label{StandardParameters}
 \begin{center} 
  \begin{tabular}{ll}
  \hline
  Parameter & Standard Value \\
  \hline
  $n_{\rmn{H}}$ & $10^3~\rmn{cm^{-3}}$ \\
  $\chi$ & 1~Draine \\
  $\zeta$ & $1.3\times10^{-17}~\rmn{s^{-1}}$ \\
  $Z$ & $Z_\odot$ \\
  $v_{\rmn{turb}}$ & $1~\rmn{km\,s^{-1}}$ \\
  $t$ & 1~Gyr \\
  \hline
  \end{tabular}
 \end{center}
\end{table}

Extinction within the cloud is calculated assuming an average grain size of $0.1\,\rmn{\umu m}$, albedo of $0.7$ and mean photon scattering by grains of $g=0.9$ ($g=\langle\cos\theta\rangle$). The dust-to-gas mass ratio is assumed to scale linearly with metallicity and takes a standard value of $10^{-2}$ at Solar metallicity ($Z_\odot$). The visual extinction ($A_V$) and dust optical depth ($\tau_v$) are related by:

\begin{equation}\label{Extinction}
 A_V = 1.086\,\tau_v
\end{equation}

The incident FUV radiation is characterised by the standard Draine field \citep{Draine1978}, scaled by a free parameter $\chi$. Attenuation of the radiation field strength with cloud depth is given by:

\begin{equation}\label{Attenuation}
 \chi = \chi_0\exp(-kA_V)
\end{equation}
where $k=1.38$ and $\chi_0$ is the unattenuated field strength at the surface of the cloud (in units of the Draine field).

Self-shielding of H$_2$ against photodissociation is calculated using the single line approximation of \citet*{Federman1979}. The self-shielding factors of \citet{vanDishoeck1988} are adopted for CO. Neutral carbon shielding includes contributions from H$_2$, CO and \mbox{C\,{\sc i}}, using the treatment of \citet{Kamp2000}.

The gas temperature is determined at each depth point using an iterative process to balance the total heating and cooling rates. Photoejection of electrons from dust grains \citep{Tielens1985} and PAHs \citep{Bakes1994}, and FUV pumping and photodissociation of H$_2$ molecules \citep{Hollenbach1979} are the dominant heating mechanisms near the cloud surface, whilst cosmic ray heating \citep{Tielens1985} and exothermic reactions become significant deeper into the cloud. Heating by \mbox{C\,{\sc i}} photoionization \citep{Kamp2001}, H$_2$ formation \citep{Kamp2001}, gas-grain collisions \citep{Burke1983} and turbulence \citep{Black1987} are also included in the model.

The gas is mainly cooled through emission from collisionally excited atoms and molecules and by interactions with the cooler dust grains. Emission from the [\mbox{O\,{\sc i}}], [\mbox{C\,{\sc i}}] and [\mbox{C\,{\sc ii}}] fine structure lines and CO rotational lines (up to $J=11\to10$) are calculated at each depth point in the code using the escape probability method of \citet*{deJong1980} and non-LTE level populations determined using the updated collisional rate coefficients from the recent PDR benchmarking \citep{Rollig2006}. The contribution to the total cooling rate by H$_2$ emission is calculated using data from \citet*{Martin1996}. H$\alpha$ and \mbox{O\,{\sc i}} 6300\AA\ fine structure line emission are also considered.

Variation in metallicity is accounted for by assuming that the dust and PAH photoelectric heating rates are directly proportional to $Z$, in addition to the scaling of the elemental abundances, H$_2$ formation rate, dust-to-gas mass ratio and grain number density.

The integrated intensity of the CO(1--0) line along the line of sight towards the cloud is then:

\begin{equation}\label{IntegratedIntensity}
 I = \frac{1}{2\upi} \int\Lambda(z)\,\rmn{d}z\qquad[\rmn{erg\,s^{-1}\,cm^{-2}\,sr^{-1}}] 
\end{equation}
where $\Lambda(z)$ is the CO(1--0) line emissivity at depth $z$ into the cloud ($\rmn{erg\,s^{-1}\,cm^{-3}}$) and the factor of $1/2\upi$ accounts for the fact that photons only emerge from the outer edge of the cloud, rather than over the full $4\upi$ steradians.

The theoretical antenna temperature, integrated over the line frequency, is given by:

\begin{equation}\label{AntennaTemperature}
 \int T_{\rmn{A}}\,\rmn{d}v = \frac{c^3}{2k_B\nu^3}\ I\qquad[\rmn{K\,km\,s^{-1}}]
\end{equation}

The value of $X$ is therefore calculated at each depth point into the PDR using equation~\ref{XFactor}.

\subsection{The Meudon code}\label{Meudon}
The Meudon PDR code self-consistently computes the steady-state chemical and thermal structure of the cloud, in addition to the emission strengths of the relevant cooling lines. A new version of the Meudon code is now available \citep[for details, see][]{LePetit2006}, which includes a detailed treatment of the ultraviolet radiative transfer and the extension to finite plane-parallel geometry. However, the present benchmark calculations have been performed using an older version of the code \citep{Abgrall1992,LeBourlot1993b,Flower1994}, in which self-shielding of H$_2$ and CO are computed using the FGK approximation \citep{Federman1979} and do not include the overlap between H$_2$ and CO predissociating lines. This has been demonstrated to be a reasonable approximation by \citet{Lee1996}.

\section{Benchmarking}\label{Benchmarking}
In order to assess the reliability of the model predictions in this work, the $X$ factor data produced by the \textsc{ucl\_pdr} code have been benchmarked against the model results of the Meudon PDR code to check that the same trends are found in both. The Meudon code treats the microphysical processes within the cloud in more detail than the \textsc{ucl\_pdr} code, but is more computationally intensive and less suited to examining large regions of parameter space. Therefore, to test the agreement between the two codes, we have selected six key models (listed in Table~\ref{BenchmarkTests}) that are representative of the parameter space being examined.

\begin{table}
 \caption{Benchmark model parameters. When not specified, the standard parameter values listed in Table~\ref{StandardParameters} were used.}
 \label{BenchmarkTests}
 \begin{center} 
  \begin{tabular}{ll}
  \hline
  Model & Parameter \\
  \hline
  A & Standard Values \\
  B & $n_{\rmn{H}}=500~\rmn{cm^{-3}}$ \\
  C & $\chi=10^3$~Draine \\
  D & $\zeta=1.3\times10^{-14}~\rmn{s^{-1}}$ \\
  E & $Z=10^{-2}\times Z_\odot$ \\
  F & $v_{\rmn{turb}}=3~\rmn{km\,s^{-1}}$ \\
  \hline
  \end{tabular}
 \end{center}
\end{table}

\begin{figure*}
 \begin{center}
  \includegraphics[width=0.95\textwidth]{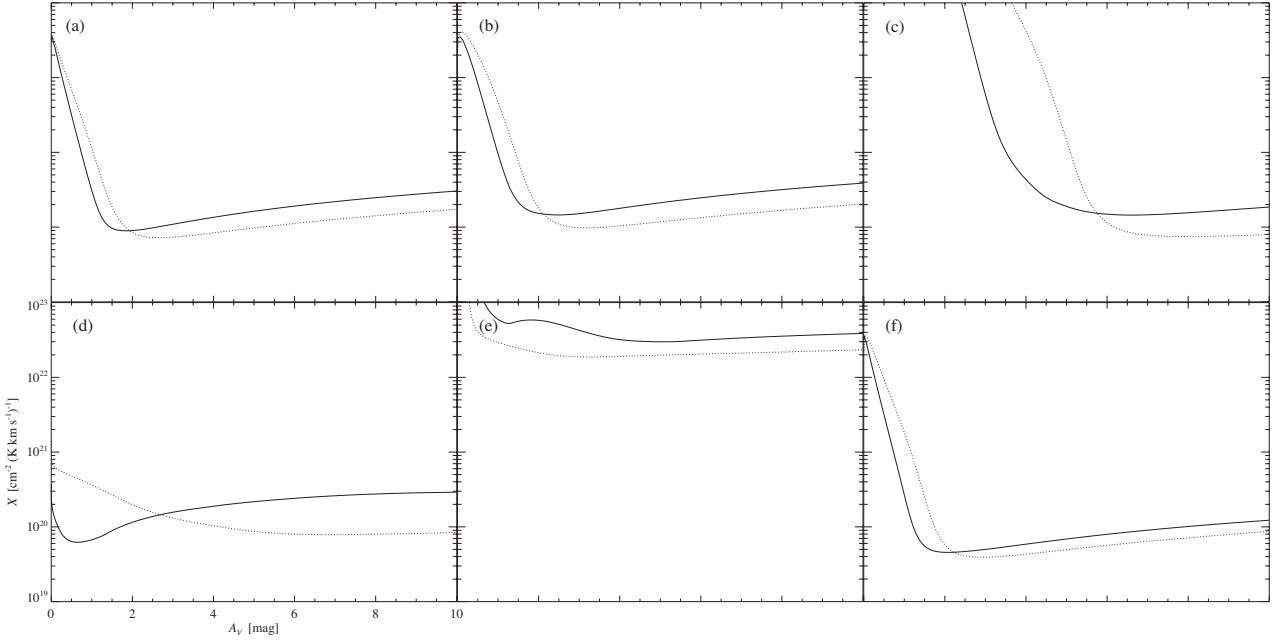}
 \end{center}
 \caption{Benchmark model comparisons. $X$ factor as a function of $A_V$, as calculated by the \textsc{ucl\_pdr} (solid line) and Meudon (dotted line) codes, is plotted for models A (standard parameters; panel a), B ($n_{\rmn{H}}=500~\rmn{cm^{-3}}$; panel b), C ($\chi=10^3$~Draine; panel c), D ($\zeta=10^3$; panel d), E ($Z=10^{-2}$; panel e) and F ($v_{\rmn{turb}}=3~\rmn{km\,s^{-1}}$; panel f).}
 \label{BenchmarkResults}
\end{figure*}

The same species set and reaction network have been used in both codes as part of the benchmarking process. The grain properties, extinction law and metallicity-dependence are also common to both codes, so that the results can be directly compared. Since the Meudon code computes steady-state chemistry, a cloud age of 1~Gyr has been used in the \textsc{ucl\_pdr} code to approach chemical equilibrium.

Fig.~\ref{BenchmarkResults} shows the results of the benchmarking tests. It must be stressed that exact agreement between the two codes is not expected, since their treatment of the physical processes is very different in some cases. Overall, agreement is good between the two codes, with $X$ profile minima generally differing by less than a factor of 2 across the visual extinction range considered here (see Table~\ref{BenchmarkXvalues}). The test models probe the influence of each parameter considered in this paper and both codes predict the same trend for the $X$ factor variation in each case. However, the Meudon code produces $X$ factor depth profiles with minima that are consistently lower and deeper into the cloud than those calculated by the \textsc{ucl\_pdr} code. This can be attributed to the different treatments of the self-shielding used by the two codes in the attenuation of the CO photodissociation and \mbox{C\,{\sc i}} photoionization rates (see Sections~\ref{UCLPDR} and \ref{Meudon} for details). The Meudon models also consider the increased photodissociation rate of H$_2$ molecules in rotationally excited states, an effect neglected in the \textsc{ucl\_pdr} code. The differences caused by the individual self-shielding treatments and the additional photodissociation of excited H$_2$ included in the Meudon code are most apparent in benchmark model C ($\chi=10^3$~Draine; panel c in Fig.~\ref{BenchmarkResults}) where the FUV radiation is strong and the gas temperatures are high. These differences should be preserved, since they reflect the choice of treatment for this physical process and give an indication of the uncertainty inherent in calculations of theoretical $X$ factors. We regard the agreement between the codes as satisfactory and the exploration of parameter space reported below is made using the \textsc{ucl\_pdr} code.

\begin{table}
 \caption{Minimum $X$ values obtained for the six benchmark models, as calculated by the \textsc{ucl\_pdr} and Meudon codes.}
 \label{BenchmarkXvalues}
 \begin{center} 
  \begin{tabular}{l . .}
  \hline
  \multicolumn{3}{c}{$X_{\rmn{min}}$ $[10^{20}~\rmn{cm^{-2}\,(K\,km\,s^{-1})^{-1}}]$} \\
  Model & \multicolumn{1}{c}{\textsc{ucl\_pdr}} & \multicolumn{1}{c}{Meudon} \\
  \hline
  A & 0.9 & 0.7 \\
  B & 1.5 & 1.0 \\
  C & 1.4 & 0.7 \\
  D & 0.6 & 0.8 \\
  E & 300.0 & 190.0 \\
  F & 0.5 & 0.4 \\
  \hline
  \end{tabular}
 \end{center}
\end{table}

\section{Model results and sensitivity of $\bmath{X}$ to parameter variations}\label{Results}

We now examine the variation of the $X$ factor under different physical conditions. The influence of changing the number density of hydrogen nuclei ($n_{\rmn{H}}$), incident radiation field strength ($\chi$), cosmic ray ionization rate ($\zeta$), cloud age ($t$), metallicity ($Z$) and turbulent velocity ($v_{\rmn{turb}}$) are considered individually. With the exception of the parameter being examined, the standard values listed in Table~\ref{StandardParameters} are adopted for each model. Plots of $X$ as a function of visual extinction ($A_V$) into the cloud are used to illustrate these trends for each parameter.

\begin{figure}
 \includegraphics[width=84mm]{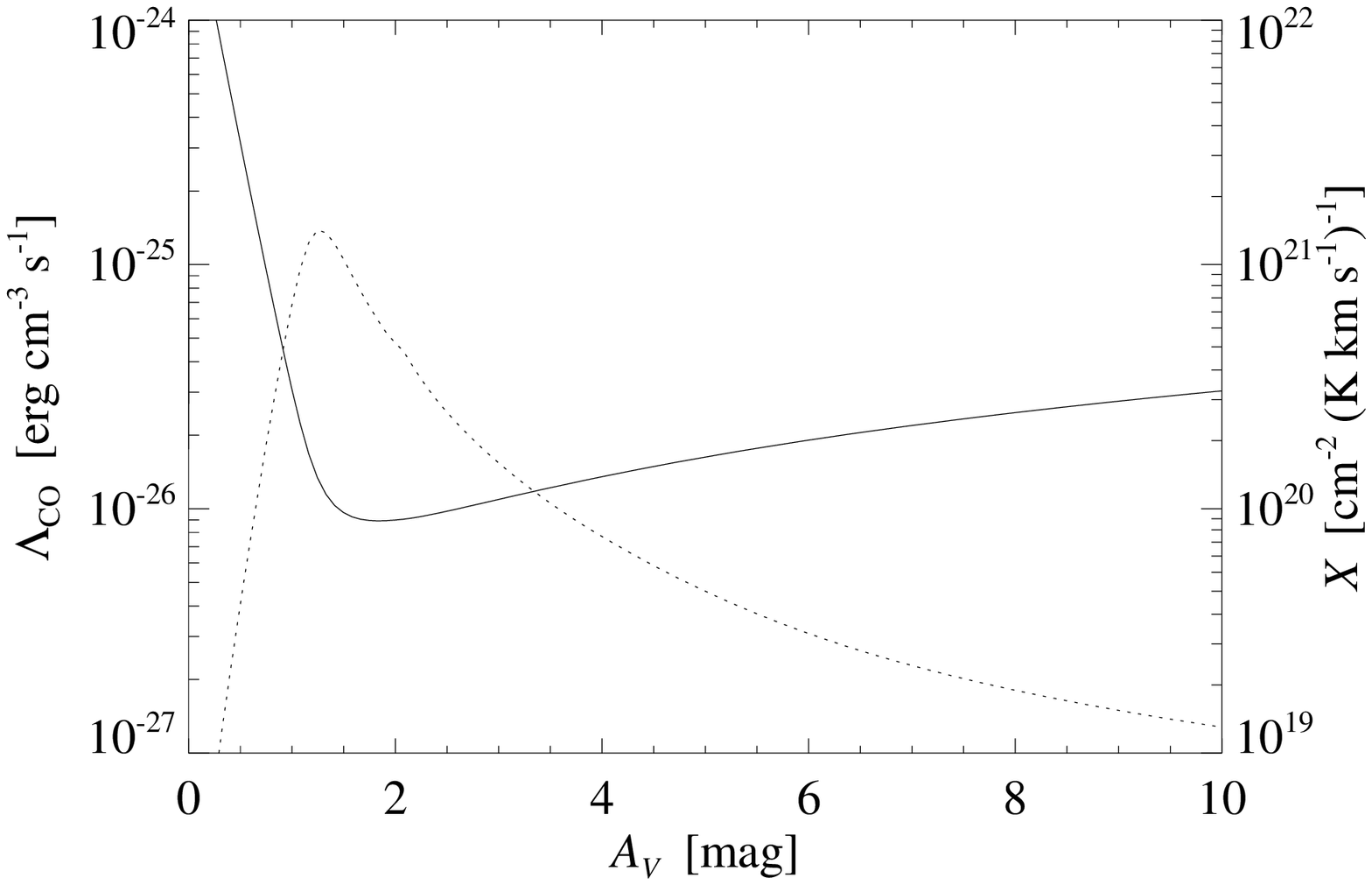}
 \includegraphics[width=84mm]{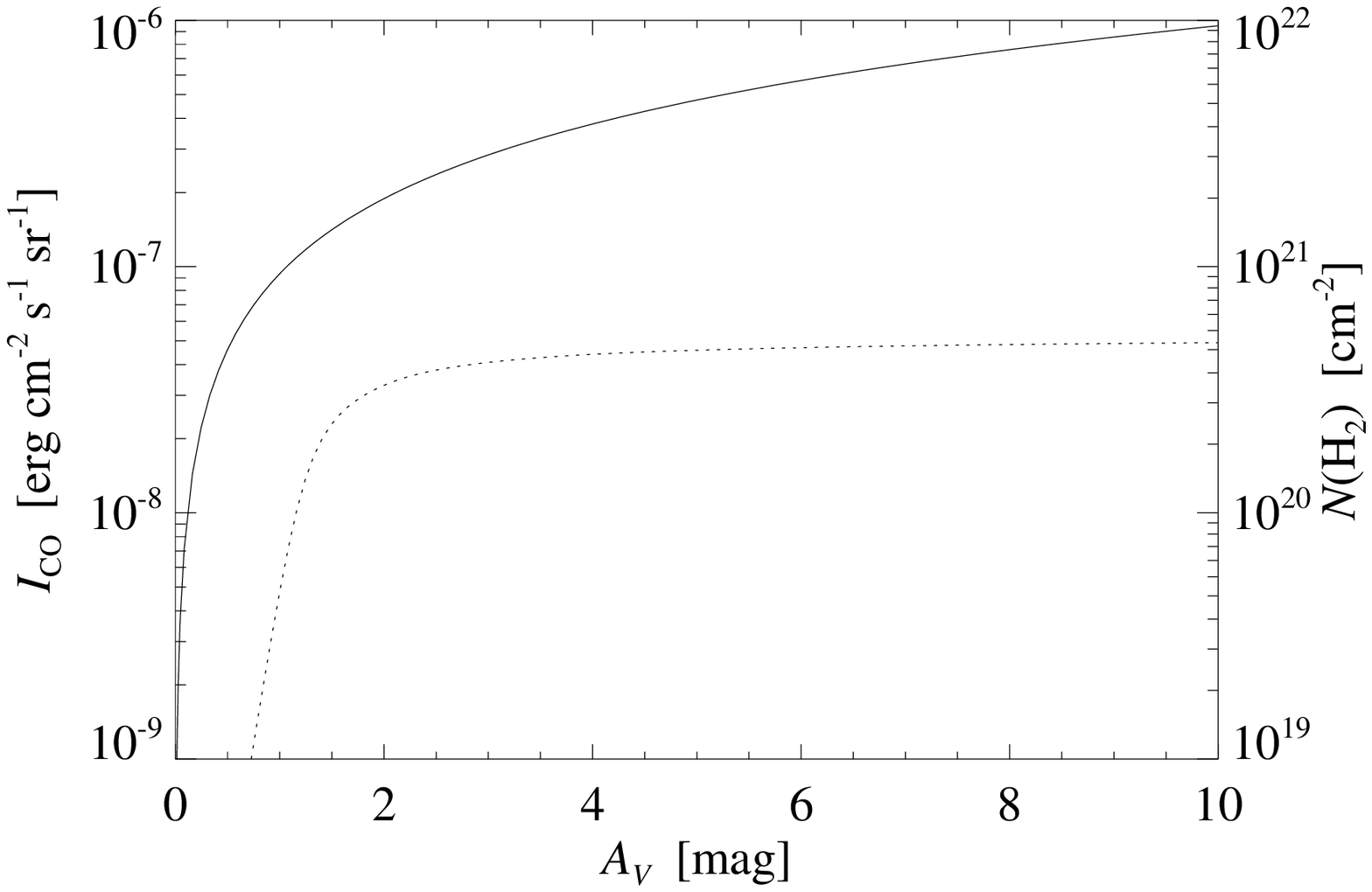}
 \caption{\textit{Top}: an $X$ factor profile calculated for the standard parameter values listed in Table~\ref{StandardParameters}. CO(1--0) emissivity ($\Lambda$; dotted line) and $X$ (solid line) are plotted as a function of $A_V$ into the cloud. \textit{Bottom}: CO(1--0) integrated intensity (dotted line) and H$_2$ column density (solid line) as a function of $A_V$ into the cloud.}
 \label{EmissionPlot}
\end{figure}

The characteristic depth profile of the $X$ factor has three main features. Near the outer edge of the cloud, where the $\rmn{H/H_2}$ transition occurs and H$_2$ is present without CO, $X$ is large. At higher visual extinction, while all hydrogen is now in molecular form, CO self-shielding becomes sufficiently strong to prevent photodissociation, allowing its abundance and emission strength to rise sharply. This causes $X$ to drop, reaching a minimum value in the region where CO emission is strongest. As the CO line becomes optically thick with increasing depth, local emission gradually declines and $X$ begins to rise slowly again. The minimum of the depth profile therefore represents the value of $X$ that would be observed for that particular environment, since it corresponds to the peak emission of the CO line (see Fig.~\ref{EmissionPlot}), with little contribution from deeper within the cloud. Under most conditions, $X$ profile minima tend to occur in the region $A_V<2$~mag where freeze-out of atoms and molecules should not take place, therefore our decision to exclude it from the chemistry seems reasonable.

The lower panel of Fig.~\ref{EmissionPlot} demonstrates that once the CO line has become optically thick, its integrated intensity reaches a constant value and is no longer sensitive to the monotonically increasing H$_2$ column density deeper into the cloud. Therefore, in the case of a slab of uniform density, the $X$ factor only provides a lower limit for $N(\rmn{H_2})$. However, in reality, there is good evidence that at least some interstellar clouds are clumpy \citep*[see, for example,][]{Peng1998,Morata2005}. In this case, the CO emission arises from an ensemble of clumps, each effectively optically thin. The total H$_2$ column density is then recovered using the $X$ factor method if clouds are clumpy on scales of $A_V\sim1$~mag, as has been suggested previously \citep[e.g.][]{Taylor1993,Wolfire1993}.

For our choice of standard physical parameters adopted in Fig.~\ref{EmissionPlot}, the derived value of $X$ would be about 1, rather than the canonical value for the Milky Way of 2, in units of 10$^{20}~\rmn{cm^{-2}\,(K\,km\,s^{-1})^{-1}}$. We note, however, that we could obtain $X=2$ with a variety of plausible parameter combinations, e.g., by setting the cloud number density to $400~\rmn{cm^{-3}}$, or the cloud age to $\sim$1~Myr, or the turbulent velocity to $0.3~\rmn{km\,s^{-1}}$. This degeneracy in reproducing the observed value therefore shows that $X$ cannot be used to infer the physical conditions and, conversely, that using the canonical value in any situation without further consideration can be dangerous.

\begin{figure}
 \includegraphics[width=84mm]{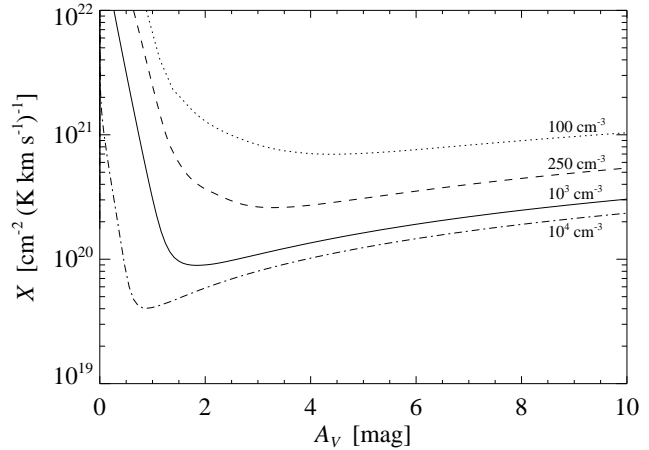}
 \caption{$X$ versus $A_V$ profiles for varying number density of hydrogen nuclei within the cloud ($n_{\rmn{H}}$ $\rmn{cm^{-3}}$).}
 \label{DensityPlot}
\end{figure}

\subsection{Gas density}
Diffuse clouds containing reasonable quantities of molecular gas have hydrogen nuclei number densities $n_{\rmn{H}} \la 100~\rmn{cm^{-3}}$, whilst giant molecular clouds have densities ranging from an average value of $\sim$$200~\rmn{cm^{-3}}$ up to $10^6~\rmn{cm^{-3}}$ in their cores \citep[see, e.g.,][ and references therein]{Scoville1987,Gusten1989,Dyson1997}. Therefore, we consider the sensitivity of $X$ to variations in density within the range $10^2 \le n_{\rmn{H}} \le 10^6~\rmn{cm^{-3}}$. For the purposes of comparison, we have adopted the simplifying assumption of constant density across the PDR in these models. The variation of $X$ with visual extinction for densities from $10^2$ to $10^4~\rmn{cm^{-3}}$ is shown in Fig.~\ref{DensityPlot}.

As the density increases from the lowest values considered, the minimum value of $X$ drops and the profile minimum gets sharper and shifts to lower $A_V$. The effective shielding of CO from photodissociation increases with gas density and allows its abundance to reach saturation at lower $A_V$ values, resulting in stronger emission and therefore reducing the value of $X$, which is inversely proportional to CO emission strength. At higher densities, the CO(1--0) line rapidly becomes optically thick with increasing depth, causing emission to occur within a thin region of the cloud and narrowing the minimum of the $X$ versus $A_V$ profile. At densities above $10^4~\rmn{cm^{-3}}$ (not shown) the sharpening and deepening of the $X$ profile minimum becomes more exaggerated.

\begin{figure}
 \includegraphics[width=84mm]{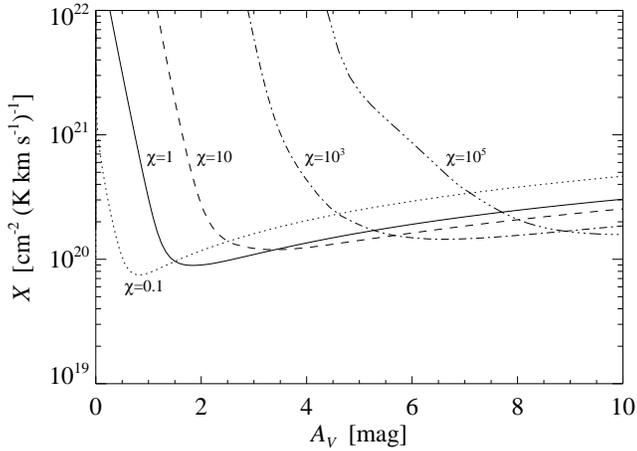}
 \caption{$X$ versus $A_V$ profiles for varying incident FUV radiation field strength ($\chi$ Draine).}
 \label{RadiationPlot}
\end{figure}

\subsection{Radiation field strength}
Radiation field strengths incident upon gas clouds can vary from a few times the standard interstellar Draine field up to $\chi\sim10^7$~Draine in regions of intense star formation \citep[see][and references therein]{Hollenbach1999}. We adopt a range of $0.1 \le \chi \le 10^5$~Draine for consideration in this study. Fig.~\ref{RadiationPlot} shows $X$ versus $A_V$ profiles for radiation field strengths within that range.

The minimum of the profile increases, broadens and moves deeper into the cloud with increasing radiation strength. The C$^+$/C/CO transition region occurs at higher values of $A_V$ as the radiation penetrates deeper into the cloud and the rise in CO abundance is more gradual under stronger radiation, as shielding becomes less effective. This leads to a broader emission feature, causing the wider profiles seen in Fig.~\ref{RadiationPlot} as the radiation intensity increases.

\begin{figure}
 \includegraphics[width=84mm]{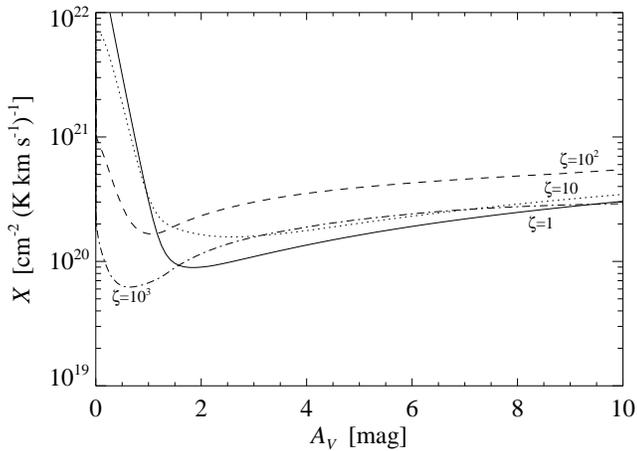}
 \caption{$X$ versus $A_V$ profiles for varying cosmic ray ionization rate ($\zeta$ $1.3\times10^{-17}~\rmn{s^{-1}}$).}
 \label{CRPPlot}
\end{figure}

\subsection{Cosmic ray ionization rate}
The cosmic ray flux is known to vary by over an order of magnitude. Magnetic field lines can channel cosmic rays away from dense molecular cores; alternatively, the flux of particles in starburst regions can be many times higher than the canonical rate \citep{Schilke1993}. We therefore consider a range of $0.1 \le \zeta \le 10^3$ times our adopted canonical rate of $1.3\times10^{-17}~\rmn{s^{-1}}$.

Increasing the flux of ionizing particles through the cloud produces a complex behaviour in the $X$ versus $A_V$ profiles (see Fig.~\ref{CRPPlot}). As the ionization rate increases from the standard value (see Table~\ref{StandardParameters}), CO is initially destroyed more effectively through reactions with He$^+$ and its abundance drops, causing an upward shift in the $X$ profile. This is the case for the $\zeta=10$ and $\zeta=10^2$ models. However, the corresponding increase in the cosmic ray heating rate causes a rise in gas temperature, becoming the dominant heating mechanism at $A_V>1$~mag for $\zeta=10^2$ and at all depths for $\zeta=10^3$. This promotes CO formation and emission, and counters, in part, the increased destruction rate. A hump emerges in the CO abundance and emission profiles (not shown) across the region $0<A_V<1.5$~mag for ionization rates above $10^2$, produced by the strong cosmic ray heating and effective [\mbox{O\,{\sc i}}] cooling that peaks at $A_V\sim1.5$~mag. This results in $X$ profile minima that are lower than in the standard model and nearer the cloud surface. Thus, a slight rise in ionization rate shifts the $X$ profile minimum upwards, whilst a further increase in ionization rate leads to the $X$ minimum dropping and moving towards the cloud surface. In models with $\zeta$ below the standard value (not shown), there is no appreciable change in the $X$ profile shape.

\begin{figure}
 \includegraphics[width=84mm]{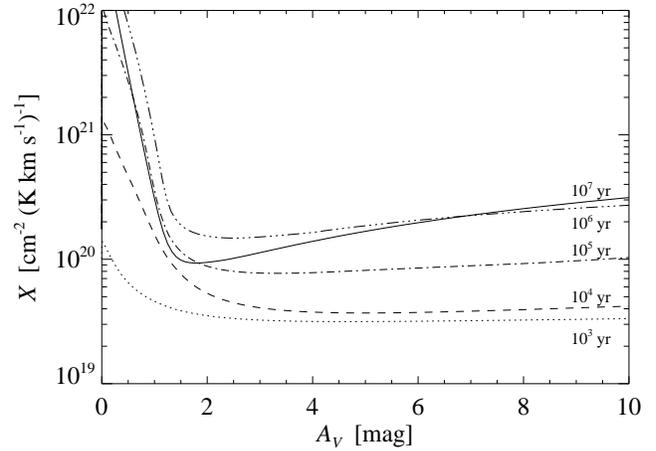}
 \caption{$X$ versus $A_V$ profiles for varying cloud age ($t$ yr).}
 \label{TimePlot}
\end{figure}

\subsection{Time-dependence}\label{Time}
The evolution of a cloud consisting initially of atomic species produces profiles that change from low, very broad minima to the usual, sharper minimum as the chemistry approaches equilibrium (see Fig.~\ref{TimePlot}). The formation of molecular hydrogen represents the significant time-scale in the chemical evolution, since the formation of CO proceeds fairly rapidly once H$_2$ is dominant. At early times, hydrogen is mainly in atomic form and the CO abundance is low, reaching a value of $\sim$$10^{-8}$ for depths beyond the point at which photodissociation becomes negligible. The low CO abundance and higher gas temperature mean that the CO(1--0) line does not become saturated and emission extends deep into the cloud, producing profiles with very broad minima. Once CO reaches abundances $>$$10^{-5}$ (which occurs at an age of $\sim$1~Myr for the standard parameter values) emission from the $J=1\to0$ line becomes optically thick, producing an emission peak at $A_V\sim1.2$~mag and the characteristic sharp minimum in the $X$ profile. For the standard parameters considered in this model (see Table~\ref{StandardParameters}) there is no noticeable change in the $X$ profile for cloud ages greater than 10~Myr, even though the chemistry does not reach steady-state until 100~Myr. This is because emission from the optically thick CO(1--0) line is no longer sensitive to the change in CO abundance at depths beyond the emission peak.

\subsection{Metallicity}\label{Metallicity}
\begin{figure}
 \includegraphics[width=84mm]{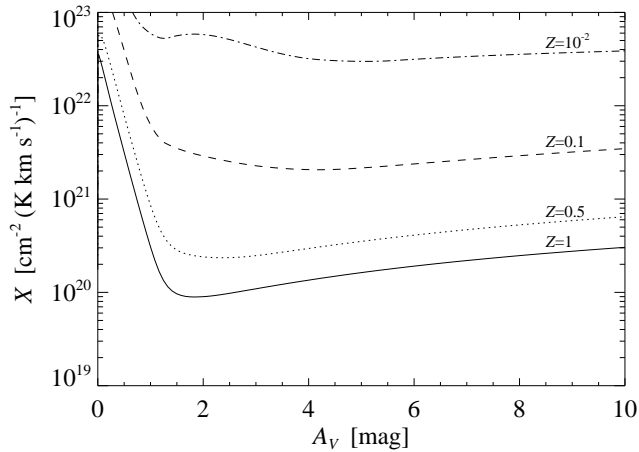}
 \caption{$X$ versus $A_V$ profiles for varying metallicity ($Z/Z_\odot$).}
 \label{MetallicityPlot}
\end{figure}

Metallicities a factor of 10 below Solar have been observed in Local Group galaxies, including the Small Magellanic Cloud \citep{Lequeux1979}, and more distant galaxies are believed to have metallicities as low as 10$^{-2}$. We adopt a metallicity range of $10^{-2} \le Z/Z_\odot \le 1$ for this study. The dependence of $X$ upon metallicity has been the subject of previous observational studies \citep[e.g.][]{Wilson1995,Arimoto1996} that find a similar trend to that shown in Fig.~\ref{MetallicityPlot}, namely, that the value of $X$ is generally increased when metallicity is less than Solar. This is simply due to the reduced abundance of CO that can form, resulting in a matching drop in CO(1--0) emission. There are, however, changes in the profile shape as the metallicity decreases. The reduction in metallicity also inhibits H$_2$ formation due to depletion of the grain surface area available for catalysis and this has implications for the time-scales required to reach chemical equilibrium. For the $Z=10^{-2}$ model, the abundance of H$_2$ does not reach steady-state until 1~Gyr, a time-scale that is unlikely to be reached under real conditions, where turbulence and dynamical evolution prevent clouds remaining quiescent for such long periods. Low metallicity environments can therefore be far from chemical equilibrium, in which case the appropriate value of $X$ can be significantly different from that derived under the assumption of steady-state.

\subsection{Turbulent velocity}\label{Turbulence}
The degree of turbulence within a cloud can be parametrized by its contribution to the observed random velocities. Turbulent velocities can be close to zero in quiescent clouds and $5~\rmn{km\,s^{-1}}$ or more in particularly active regions \citep[e.g.][]{Williams1995,Brunt2002}. We consider a turbulent velocity range $0.5 \le v_{\rmn{turb}} \le 3~\rmn{km\,s^{-1}}$ in this work.

\begin{figure}
 \includegraphics[width=84mm]{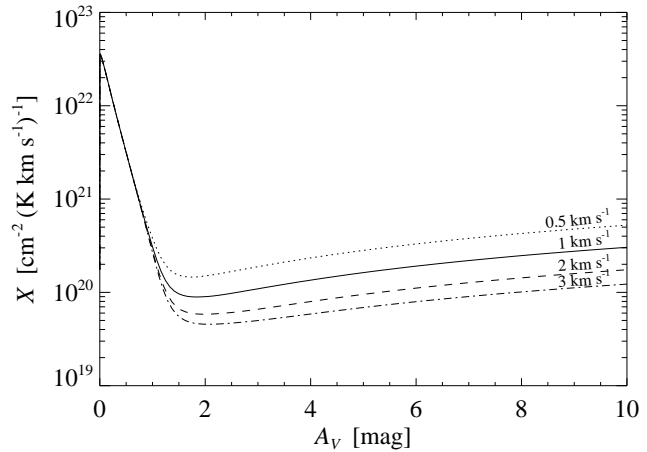}
 \caption{$X$ versus $A_V$ profiles for varying turbulent velocity within the cloud ($v_{\rmn{turb}}$ $\rmn{km\,s^{-1}}$).}
 \label{TurbulencePlot}
\end{figure}

An increase in the turbulent velocity of the cloud serves to decrease the value of $X$ at each depth point whilst retaining the characteristic profile shape (see Fig.~\ref{TurbulencePlot}). The stronger turbulent broadening of emission lines leads to lower opacities, causing a rise in CO emission, and a corresponding drop in the value of $X$.

\section{Conclusions}\label{Conclusions}
This work presents a thorough investigation of the properties and behaviour of the $X$ factor under varying physical parameters. Such an understanding of the variation in $X$ is vital when using the CO-to-H$_2$ method to estimate molecular gas mass in regions that are markedly different to the local interstellar medium. We have demonstrated that the conversion factor can vary by over an order of magnitude under the influence of some physical parameters, in particular the density, age and metallicity of the region. The benchmarking of the PDR code provides support for these predictions, whilst highlighting the inherent uncertainties in theoretical models of this kind. To summarize the trends discussed in this work:

\begin{enumerate}
\item Increasing density causes the minimum value of $X$ to drop and shift to lower $A_V$.
\item Stronger radiation fields slightly raise the $X$ profile minimum and push it deeper into the cloud.
\item A slight increase in the cosmic ray ionization rate will raise the $X$ minimum, but ionization rates over 100 times the standard value result in the minimum dropping and appearing nearer to the cloud surface.
\item Turbulent velocity variation does not significantly alter the shape or position of the $X$ profile minimum; increasing turbulent velocity produces a slight drop in the $X$ value.
\item Young, predominantly atomic regions produce broad, low $X$ profile minima; the characteristic $X$ profile shape only appears once a cloud has undergone considerable chemical evolution, corresponding to cloud ages above 1~Myr.
\item Decreasing metallicity has a strong effect on the $X$ profile, increasing the minimum by over 2 orders of magnitude above the canonical value in the $Z=10^{-2}$ case.
\item The issue of time-dependence in low metallicity environments is particularly important, since the reduced H$_2$ formation rate means that the chemistry does not reach steady-state for cloud ages below 100~Myr.
\end{enumerate}

Conditions that produce extended CO emission deep into the cloud (such as low density, strong radiation fields or low metallicity) serve to broaden the $X$ profile minimum, and the CO emission reliably traces the H$_2$ column density to higher $A_V$ in such cases. Since interstellar clouds are likely to be clumpy on scales significantly less than the 10 mag of extinction considered in these models, the CO emission in each clump is likely to be optically thin for most of the parameters considered here. Conditions producing extended emission allow for the presence of larger clump sizes before the reliability of the $X$ factor as a tracer of total H$_2$ column density breaks down.

In future work, we will derive a set of $X$ factor values appropriate for specific extragalactic environments and compare them to observational studies of the $X$ factor in similar regions. This will allow the accuracy of theoretical modelling of $X$ to be evaluated and we will present a table of conversion factors that may be used by observers to obtain more reliable estimates for molecular gas masses, along with limitations for their use.

\section*{Acknowledgments}
TAB is supported by a PPARC studentship. SV acknowledges individual financial support from a PPARC Advanced Fellowship. We thank the referee for constructive comments which helped to improve an earlier draft of this paper.

\bsp

\label{lastpage}

\end{document}